\documentclass[slac_one]{revtex4}
\usepackage{graphicx}
\usepackage{fancyhdr}
\usepackage{subfigure}

\newcommand{\bq}{\begin{equation}}
\newcommand{\eq}{\end{equation}}

\def\MW{$\mathrm{M}_{\mathrm{W}}$}
\def\MH{$\mathrm{M}_{\mathrm{H}}$}

\def\mt{$\mathrm{m}_{\mathrm{t}}$}

\begin{document}

\title{{\small{34th International Conference on High Energy Physics (ICHEP08), Philadelphia, USA}}\\ %% Please keep this conference title here
\vspace{12pt}
Electroweak Physics with ATLAS} %% Paper title goes here


\author{Arif Akhundov} 
%for the ATLAS Collaboration}
\email{Arif.Akhoundov@cern.ch}
\affiliation{Institute of Physics, Azerbaijan Academy of Sciences, 
           370143 Baku, AZERBAIJAN}

%\classification{12.15.-y, 12.15.Lk, 13.85.-t}
%\keywords {Electroweak interactions, Electroweak radiative corrections,
%           Hadron-induced high- and super-high-energy interactions}

\begin{abstract}
 The precision measurements of electroweak parameters of the Standard Model with the ATLAS detector at LHC are reviewed.  An emphasis is put on the bridge connecting the ATLAS measurements with the SM analysis at LEP/SLC and the Tevatron.   
\end{abstract}

\maketitle

\thispagestyle{fancy}


\section{INTRODUCTION} % Section title should be in all capitals.

The Standard Model (SM) has been impressively confirmed by successful collider experiments at the particle accelerators LEP, SLC and Tevatron
\cite{LEPEWWG,LEPEWWG_SLD:06,CDFD0:08} during the last twenty years. 

In July 2008 the Large Hadron Collider (LHC) at CERN started-up to operate and will reach a total energy of 10 TeV with colliding proton beams by October 2008~\cite{Thomas:08}. 
The general purpose of the ATLAS experiment~\cite{ATLAS} is 
designed to meet the main physics goals
of the LHC: the discovery of the Higgs boson and supersymmetric particles~\cite{Anna:07, Ellis:07,Harlander:08}.

In addition to its discovery potential the LHC constitutes
a powerful tool to make precision tests of the SM~\cite{Pralavorio:05,Erler:06,Erler:07}. During the initial two years of low luminosity running from $5\times10^{31}$~cm$^{-2}$s$^{-1}$ to $10^{33}$~cm$^{-2}$s$^{-1}$ the integrated luminosity in 2008 will be 20~pb$^{-1}$ and in 2009 100~pb$^{-1}$ per LHC experiment. This will be an ideal period to continue precision measurements started at LEP/SLC and the Tevatron~\cite{Akhundov:07} thanks to the very high rate of the SM processes.  
 
\section{PRECISION FOR THE MEASUREMENTS OF THE W AND TOP QUARK MASSES}   
The masses \MW~and \mt~are the physical input parameters of the SM and
we should know them with high precision. The results of the measurements
at LEP and the Tevatron are shown in Figure 1. 
%%%%%%%%%%%%%%%%%%%%%%%%%%%%%%%%%%%%%%%%%%%%%%%%%%%%%%%%%%%%%%%%%%%%%%%%
\begin{figure}[hbt]\centering
\begin{minipage}[c]{.45\linewidth}\centering
\includegraphics[width=6.25cm]{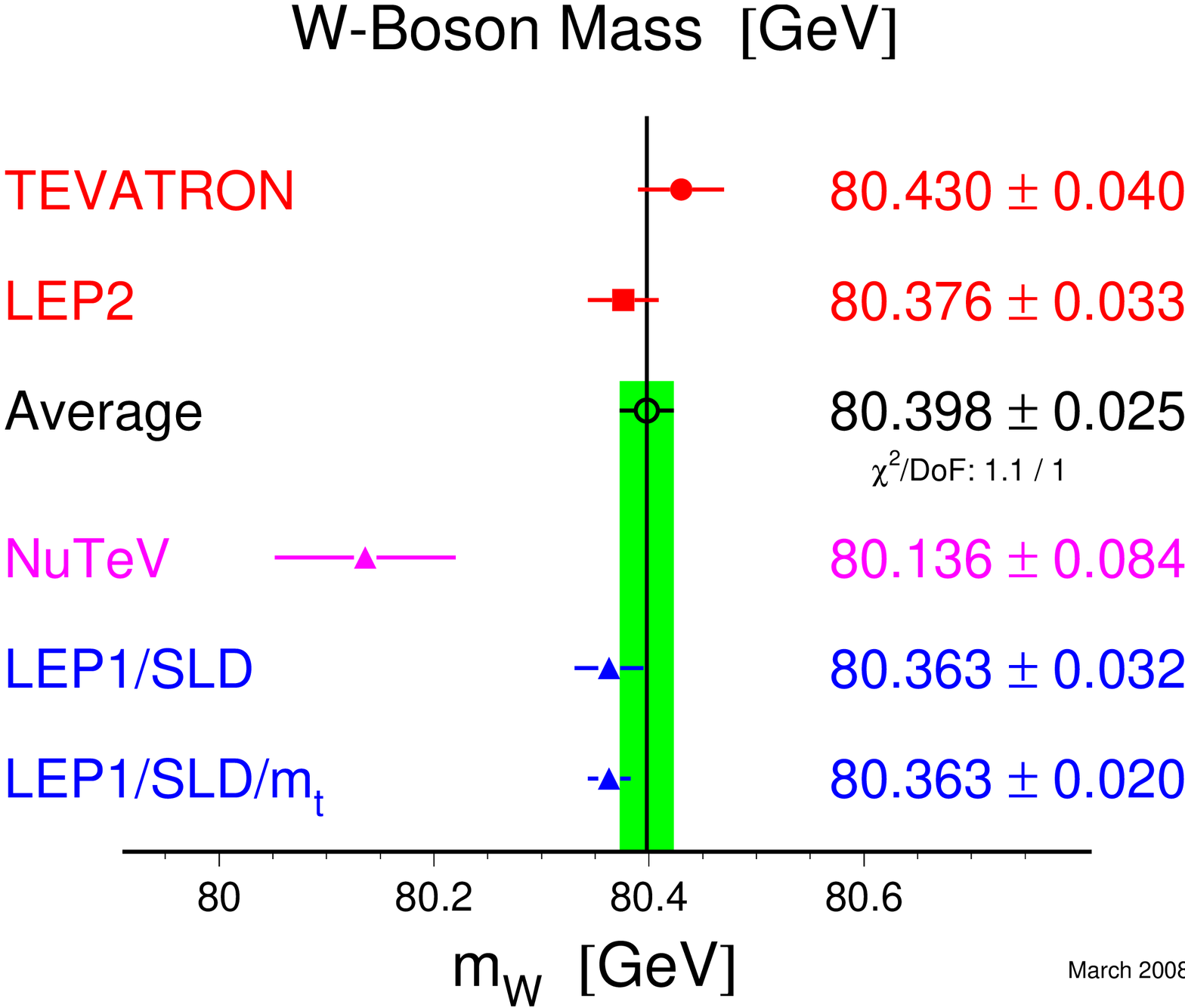}
\end{minipage}
\hskip 1cm
\begin{minipage}[c]{.45\linewidth}\centering
\includegraphics[width=6.25cm]{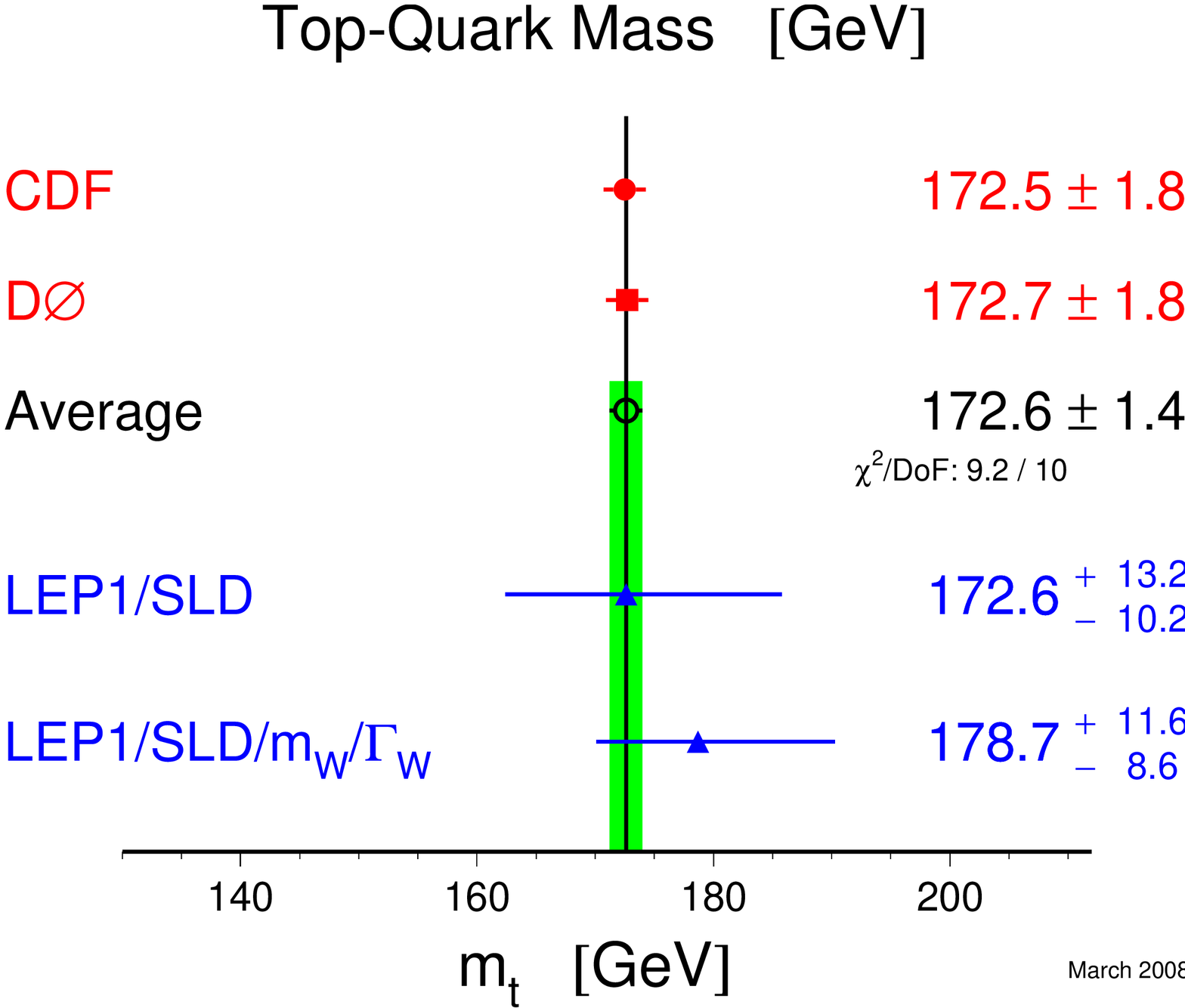}
\end{minipage}
\caption{Comparison of the direct measurements of \MW~and \mt~at LEP2 and the Tevatron with the indirect determination through electroweak radiative corrections at LEP1 and SLD~\cite{LEPEWWG,LEPEWWG_SLD:06,CDFD0:08}.} 
\end{figure}
%%%%%%%%%%%%%%%%%%%%%%%%%%%%%%%%%%%%%%%%%%%%%%%%%%%%%%%%%%%%%%%%%%%%

The higher order terms, radiative corrections or quantum corrections, contain the self-coupling of the vector bosons as well their interactions with the Higgs field and the top quark. Their calculation provides the 
theoretical basis for the electroweak precision tests~\cite{Akhundov:07}. 
Theoretical prediction of the SM on the quantum level depends on the masses \MW~and \mt~and on the as yet experimentally unknown Higgs boson 
\MH~through the virtual presence of these particles in the loops. 
The electroweak radiative corrections receive contributions from the square of the top mass~\cite{ABR1:86} and the logarithm of the Higgs mass~\cite{ABR2:86}.
As a consequence, precision data from LEP, SLC and Tevatron
\cite{LEPEWWG,LEPEWWG_SLD:06,CDFD0:08} pin down the allowed range of the mass \MW~and \mt. The Figure 2 (left) compares the information on \MW~and \mt~obtained at LEP1 and SLD with the direct measurements performed at LEP2 and the Tevatron. 

The principal method for measuring the W mass at the LHC employs the leptonic decay channels. The cross section  for $\mathrm{pp} \to \mathrm{W} + \mathrm{X}$ with $\mathrm{W} \to \mathrm{l} \nu$ ($\mathrm{l} = \mathrm{e}, \mathrm{\mu}$) is 15~nb, corresponding to $\sim 10^6$ events for the luminosity 10~fb$^{-1}$~\cite{Wielers:06}. The W mass is extracted from the distribution of the W transverse mass, $M_{\mathrm{T}}^{\mathrm{W}}$, given by
\noindent 
\begin{equation}
M_{\mathrm{T}}^{\mathrm{W}} = \sqrt{2p_{\mathrm{T}}^{\mathrm{l}} p_{\mathrm{T}}^{\nu}(1-\cos \Delta \Phi)},
\label{MWtrans}
\end{equation}
\noindent 
where $\Phi$ is the azimuthal angle between the 3-momentums of charged lepton and the system X. 

The large sample of the W events available at the LHC means that the uncertainty on the W mass, $\Delta M_{\mathrm{W}}$, arising from the statistics will be very small. The largest experimental uncertainty and the limiting factor governing the precision on the W mass is the knowledge of the 4-momentum of charged lepton which will drive a total uncertainty $\Delta M_{\mathrm{W}}$ lower than 20~MeV~\cite{Pralavorio:05}.
%%%%%%%%%%%%%%%%%%%%%%%%%%%%%%%%%%%%%%%%%%%%%%%%%%%%%%%%%%%%%%%%%%%%%%%%
\begin{figure}[hbt]\centering
\begin{minipage}[c]{.45\linewidth}\centering
\includegraphics[width=6.75cm]{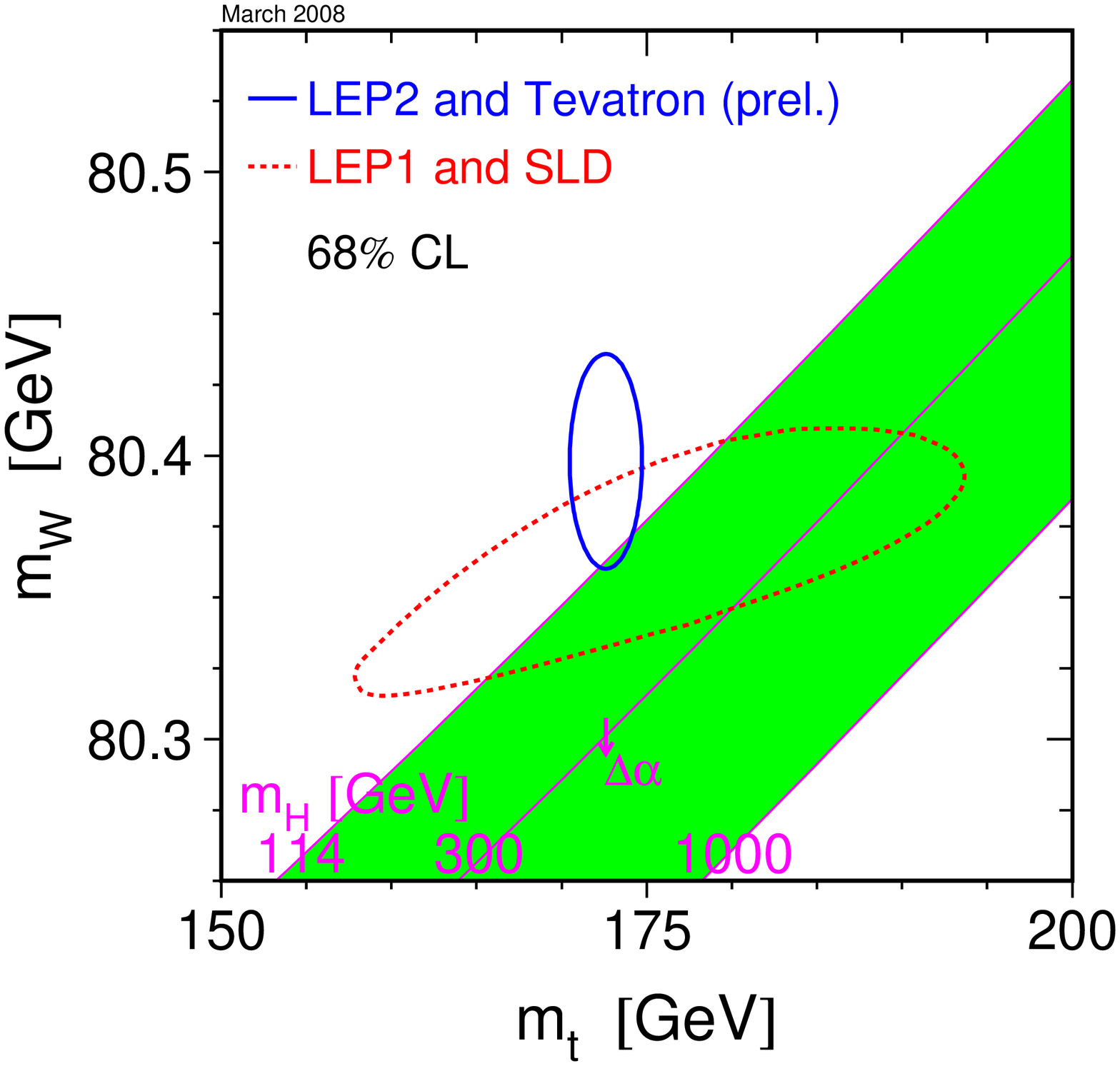}
\end{minipage}
\hskip 0.5cm
\begin{minipage}[c]{.45\linewidth}\centering
\includegraphics[width=6.75cm]{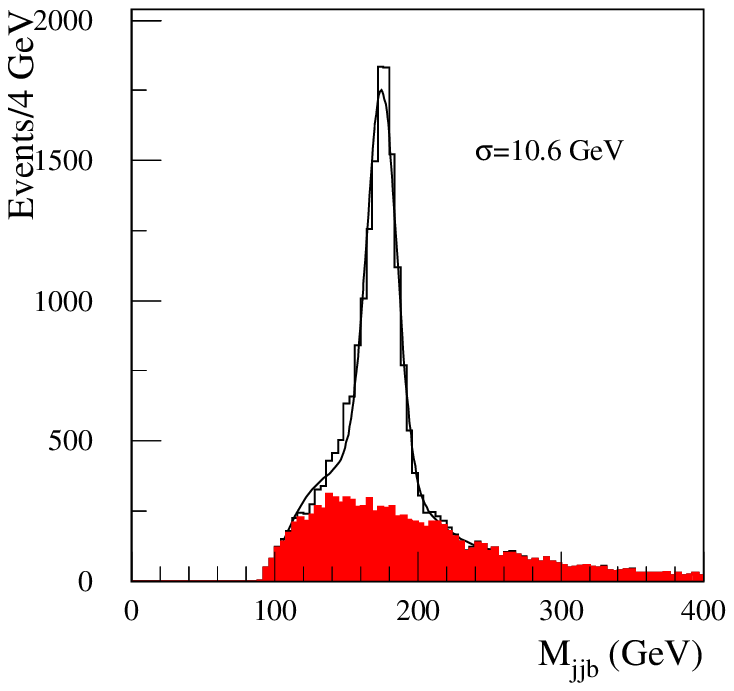}
\end{minipage}
\caption{Left: The SM relationship for the masses \MW~and
\mt~as function of \MH~\cite{LEPEWWG,LEPEWWG_SLD:06,CDFD0:08}. 
Right: Reconstructed top mass and combinatorial background (shaded area) from semileptonic $\mathrm{t\bar{t}}$ events (10 fb$^{-1}$)~\cite{Borjanovic}.} 
\end{figure}
%%%%%%%%%%%%%%%%%%%%%%%%%%%%%%%%%%%%%%%%%%%%%%%%%%%%%%%%%%%%%%%%%%%%
At LHC $\sigma(\mathrm{pp} \to \mathrm{t\bar{t}})\sim~830\pm 100$~pb which is 100 times more than at the Tevatron and a rate of $\sim 10^4$ $\mathrm{t}\bar{\mathrm{t}}$ events for the luminosity 10~fb$^{-1}$, 90\% of which are produced through $\mathrm{gg}\to \mathrm{t\bar{t}}$. The $\mathrm{t\bar{t}}$ events are categorized by the subsequent decay of the two W bosons into three types: full hadronic(65\%), semileptonic (30\%) and 
dileptonic (5\%) events. The semileptonic channel represents the most favorable route to measuring the top quark mass~\cite{Borjanovic}.
The top mass measurement proceeds by reconstructing the $\mathrm{jjb}$-system from the two jets which are not b-tagged and combining with one of the b-tagged jets. The invariant mass $\mathrm{M_{jjb}}$ distribution for the luminosity 10~fb$^{-1}$ is plotted in Figure 2 (right) together with the combinatorial background~\cite{Borjanovic}. The top quark mass is extracted from this plot by fitting to the peak using Monte Carlo samples generated with different values of \mt.

The large sample of $\mathrm{t\bar{t}}$ events available at the LHC means that the uncertainty on the top quark mass, $\Delta m_{\mathrm{t}}$, arising from the statistics also will be very small. The largest uncertainty will arise from the contribution of the final state radiation of
massive particles~\cite{Akhundov:85}. Nevertheless, ATLAS could attain a precision of  $\sim$~1~GeV on the top quark mass in two years of low luminosity running.

\section{CONCLUSIONS}

The precision measurements of the W and top quark masses with the ATLAS detector at LHC are reviewed. The anticipated uncertainty in the ATLAS measurements $\Delta \mathrm{M}_{\mathrm{W}}\sim$~15~MeV and $\Delta \mathrm{m}_{\mathrm{t}}\sim$~1~GeV would constrain the mass of the Higgs boson to within 40\%. 

% \begin{acknowledgments}
% The authors wish to thank JACoW for their guidance in preparing
% this template.
% 
% Work supported by Department of Energy contract DE-AC02-76SF00515.
% \end{acknowledgments}

\begin{acknowledgments}
 I would like to thank the organizers of the ICHEP08 for kind invitation
 to present this poster. I am grateful to A.~Di~Ciaccio for support of this
 material. I thank P.~Pralavorio for important comments and A.~Shiekh for 
 remarks.
\end{acknowledgments}
 
%\begin{thebibliography}{9}   % Use for  1-9  references
%\begin{thebibliography}{99} % Use for 10-99 references

\end{document}